\documentclass[journal=nalefd,manuscript=letter]{achemso}

\usepackage[version=3]{mhchem}

\author{Carlos Rodr\'iguez Cort\'ez}\affiliation{Sorbonne Universit\'e, CNRS, Institut des NanoSciences de Paris, INSP, F-75005 Paris, France}
\author{Moussa Mebarki}\affiliation{Groupe d’Etude de la Mati\`ere Condensée (GEMaC), Universit\'e de Versailles Saint-Quentin-en-Yvelines,
CNRS-UMR8635, Universit\'e Paris-Saclay, 78035 Versailles, France}
\author{Bruno Berini}\affiliation{Groupe d’Etude de la Mati\`ere Condensée (GEMaC), Universit\'e de Versailles Saint-Quentin-en-Yvelines,
CNRS-UMR8635, Universit\'e Paris-Saclay, 78035 Versailles, France}
\author{Dominique Demaille}\affiliation{Sorbonne Universit\'e, CNRS, Institut des NanoSciences de Paris, INSP, F-75005 Paris, France}
\author{Vincent Polewczyk}\affiliation{Groupe d’Etude de la Mati\`ere Condensée (GEMaC), Universit\'e de Versailles Saint-Quentin-en-Yvelines,
CNRS-UMR8635, Universit\'e Paris-Saclay, 78035 Versailles, France}
\author{Yunlin Zheng}\affiliation{Sorbonne Universit\'e, CNRS, Institut des NanoSciences de Paris, INSP, F-75005 Paris, France}
 \author {Pal Bhuyan}\affiliation{Sorbonne Universit\'e, CNRS, Institut des NanoSciences de Paris, INSP, F-75005 Paris, France}
\author{Boris Vodungbo}\affiliation{Sorbonne Universit\'e, CNRS, Laboratoire de Chimie Physique Mati\`ere et Rayonnement, LCPMR, Paris, France}
\author{Emmanuelle Jal}\affiliation{Sorbonne Universit\'e, CNRS, Laboratoire de Chimie Physique Mati\`ere et Rayonnement, LCPMR, Paris, France}
\author{Horia Popescu}\affiliation{Synchrotron SOLEIL, L'Orme des Merisiers, Départementale 128, 91190 Saint-Aubin, France}
\author{Nicolas Jaouen}\affiliation{Synchrotron SOLEIL, L'Orme des Merisiers, Départementale 128, 91190 Saint-Aubin, France}
\author{Yves Dumont}\affiliation{Groupe d’Etude de la Mati\`ere Condensée (GEMaC), Universit\'e de Versailles Saint-Quentin-en-Yvelines,
CNRS-UMR8635, Universit\'e Paris-Saclay, 78035 Versailles, France}
\author{Marcel Hennes}\affiliation{Sorbonne Universit\'e, CNRS, Institut des NanoSciences de Paris, INSP, F-75005 Paris, France}
\author{Franck Vidal}\email{franck.vidal@insp.jussieu.fr}
\affiliation{Sorbonne Universit\'e, CNRS, Institut des NanoSciences de Paris, INSP, F-75005 Paris, France}

\title{Synthesis and Transfer of Freestanding Strain-Engineered 
Vertically Aligned Nanocomposite Thin Films}

\begin{document}

\begin{abstract} The recent development of freestanding oxide thin films opens up exciting opportunities for the design of novel heterostructures with enhanced functionalities. Here, we explore the fabrication of membranes consisting of dense arrays of ultrathin Co$_x$Ni$_{1-x}$ nanowires epitaxially embedded in a SrTiO$_3$ matrix. Through combined x-ray absorption spectroscopy, x-ray resonant magnetic scattering, x-ray diffraction and magneto-optical experiments, we show how a SrVO$_3$-mediated lift-off process can be used to create and transfer these membranes while simultaneously preserving the structural and chemical integrity of the  self-assembled, metallic Co$_x$Ni$_{1-x}$ nanopillars. With this approach,  the large axial deformation of the embedded nanostructures is kept intact and, as a direct consequence, the magnetic properties of the nano-composite thin films remain largely unaltered after substrate removal. Our findings thus highlight a novel route for the synthesis of freestanding, strain-engineered vertically aligned heterostructures and pave the way for their future integration into spintronic and optomagnetic devices.
\end{abstract}

\begin{tocentry}
\includegraphics[width=8cm]{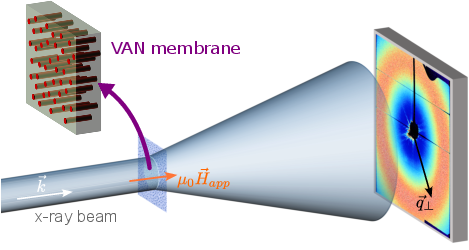}
\end{tocentry}

\noindent\textbf{keywords:} nanocomposite . membrane . epitaxy . strain-engineering . magnetic anisotropy \newline

Functional oxide membranes have emerged as a novel class of materials that has attracted increasing interest\cite{ZHA19,PESQ22, CHIA22, MAN25}. These single-crystal layers with thicknesses ranging from a few unit cells \cite{Ji2019} to hundreds of nm can be obtained by delamination of thin films grown by pulsed laser deposition (PLD) using soluble sacrifical layers \cite{lu2016synthesis, BOU20, TAKA2020}. After lift-off, they are readily transferred to other types of substrates and can be incorporated in complex hybrid or even flexible systems \cite{HUA22, SU23, GAN24, HAR23, SEG24, DEG25}. While a broad variety of thin films were already successfully removed from substrates, freestanding heterostructured layers, consisting of different phases that are epitaxially coupled to each other, have received less attention so far\cite{SHE23}. 

Within this context, self-assembled vertically aligned nanostructures (VAN) – where nm-sized columns are coherently embedded in a surrounding matrix – are particularly appealing \cite{Macmanus2010,Zheng2014,Review2018}. Because of their unique nanoarchitecture, VAN might be employed for catalysis \cite{LIP2016, LIP2021}, used in the fields of plasmonics and optical metamaterials\cite{WAN2016, WU2024}, or as building blocks for magnetoelectric \cite{ROSS2015, GAO2021} and resistive switching devices \cite{DOU2023}. A salient feature in many epitaxial VAN is their strain state governed by the lattice mismatch between the involved phases and relaxation effects at vertical heterointerfaces \cite{CHEN2013, Review2018}. In these systems, the deformations imposed by the matrix upon the wires can be used for accurate strain-mediated control of physical properties -- statically and dynamically -- as was demonstrated in earlier work\cite{LIU2012, CHE21,HEN21}. 

While the fabrication of VAN membranes appears as an attractive route to increase the portfolio of materials that can be used in novel heterostructures, to the best of our knowledge, only few reports dealing with the synthesis of freestanding composites have been published so far \cite{}. This especially holds when it comes to VAN that combine metallic and oxide phases. Huang \textit{et al}. explored the use of Sr$_3$Al$_2$O$_6$ (SAO) sacrifical layers to obtain flexible Au-TiN VAN thin film membranes and analyzed their optical properties \cite{Huang2025}. Tsai \textit{et al}. grew Au-BaTiO$_3$ VAN \cite{TSAI24} and successfully removed the membranes from the substrate. However, in these studies, the use of a noble metal reduced the risk of chemical modifications induced by the lift-off process. In composites that combine magnetic, metallic materials and an oxide matrix, the embedded nanowires consist of Fe, Co, Ni and their alloys \cite{MOH04, SHI12, Bonilla2013, SCH16, SU16, HUA17, WEN18, HEN18, WEN22, HUA2024} that can oxidize, potentially modifying their magnetic properties. In addition, the conservation (or loss) of vertical dilation -- one of the key parameters used to control the magnetic properties in VAN -- has not been analyzed in freestanding films. 

In this letter, we address this issue and report on the synthesis of magnetic metal-oxide VAN membranes where the peculiar strain state of the embedded magnetic pillars is preserved. This opens up perspectives for the design of heterostructures that cannot be obtained by conventional planar epitaxy methods. 

In previous studies on Co$_x$Ni$_{1-x}$-SrTiO$_3$ VAN, it was shown that magnetic Co$_x$Ni$_{1-x}$ nanopillars, with diameters below 5~nm and aligned along the growth direction, could be obtained via pulsed laser deposition (PLD) \cite{WEN22}. In these composites, the metallic phase grows epitaxially within the SrTiO$_3$ (STO) matrix with a cube-on-cube epitaxy. Furthermore, it was demonstrated that the metallic alloy is highly dilated along the out-of-plane direction, with tensile strains reaching several percents and giving rise to strong magnetoelastic effects \cite{WEN22}. We have thus chosen this system in order to explore the issue of metal-oxide VAN delamination. 

\begin{figure}
  \includegraphics{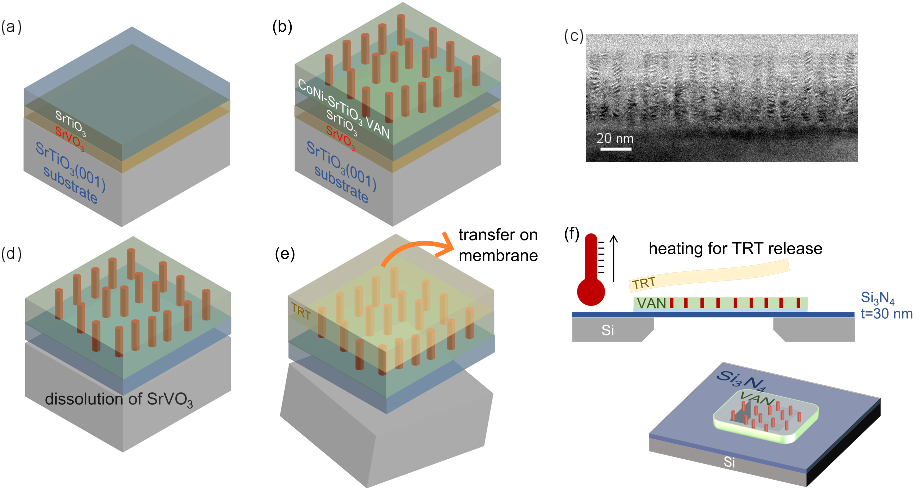}
  \caption{Growth and delamination of Co$_{0.5}$Ni$_{0.5}$-SrTiO$_3$ VAN membranes. (a) First growth step: pulsed laser deposition of homoepitaxial SrVO$_3$ and SrTiO$_3$ thin films on SrTiO$_3$(001) substrate. (b) Second growth step: sequential pulsed laser deposition of Co$_{0.5}$Ni$_{0.5}$-SrTiO$_3$ VAN on SrTiO$_3$/SrVO$_3$/SrTiO$_3$(001). (c) Cross-section transmission electron microscopy image of a Co$_{0.5}$Ni$_{0.5}$-SrTiO$_3$ VAN grown on SrTiO$_3$(001): the formation of Co$_{0.5}$Ni$_{0.5}$ nanopillars is evidenced by Moir\'e patterns. (d) Dissolution of the SrVO$_3$ sacrificial layer. (e) Removal of the substrate using a thermal release tape (TRT). (f) Transfer on Si$_3$N$_4$/Si membranes.}
  \label{fig:delam}
\end{figure}

The fabrication process is illustrated in Figure \ref{fig:delam}. Prior to deposition of the Co$_x$Ni$_{1-x}$-SrTiO$_3$ VAN, a thin 5 nm sacrificial layer of strontium vanadate SrVO$_3$ (SVO) is deposited on the STO(001) substrate \cite{BOU20}. Due to the structural compatibility (aristotype Pm$\bar{3}$m perovskite, lattice parameter 3.843 \AA \cite{FOU16}) with the STO substrate and the low lattice mismatch, this layer grows fully strained and with a cube-on-cube epitaxy, preserving the properties necessary for the subsequent deposition of a fully epitaxial VAN. On top of the SVO layer, a thin $\sim$10 nm cube-on-cube epitaxial and fully strained STO capping layer is grown  which efficiently protects the sacrificial film, and allows us to perform a transfer under ambient conditions into a second PLD chamber, where the nanocomposite thin film was eventually deposited. 

The latter was grown using a sequential scheme \cite{Bonilla2013} employing SrTiO$_3$, CoO and NiO targets as described in detail previously\cite{WEN22}. This leads to the self-assembly of nanopillars embedded in STO and aligned along the growth direction, as shown in the transmission electron microscopy images in Figure \ref{fig:delam}c, see \textbf{Supporting Information (SI)} for a more detailed description of the transmission electron microscopy results. 

\vspace{5mm}

After PLD growth of the sample, the latter is placed on a glass slide,  immersed in a pre-heated deionized (DI) water solution at 50$^\circ$C and kept at this temperature during the etching of the SVO. The sacrificial layer is completely removed in less than 24 hours (Figure \ref{fig:delam}d); this results in the release of the VAN from its substrate. After this selective etching step, the VAN (now a free-standing membrane) remains weakly bonded to the STO substrate. To remove it, the sample is carefully taken out of the DI water solution. Then, thermal release tape (TRT) is attached to the membrane, which allows us to pull it off from the STO, as shown in Figure \ref{fig:delam}e. After 24 hours of drying in ambient air, the released membrane on TRT is transferred onto a Si$_3$N$_4$ (SN) grid (Silson multielement windows with sizes of 50\,$\mu$m $\times$ 50\,$\mu$m, 100\,$\mu$m $\times$ 100\,$\mu$m and 200\,$\mu$m $\times$ 200\,$\mu$m, respectively). Subsequent annealing at 70$^\circ$C for 4 minutes, and at 90$^\circ$C for 2 minutes ensures that the VAN membrane is completely dry and adheres to the grid. In a last step, the temperature is increased to detach the TRT, completing the transfer of the VAN membrane, as shown in Figure \ref{fig:delam}f. In this work, 3 different types of freestanding VAN membranes with distinct composition of the magnetic nanopillars were eventually grown and analyzed in detail: Co$_{0.2}$Ni$_{0.8}$, Co$_{0.5}$Ni$_{0.5}$ and Co$_{0.8}$Ni$_{0.2}$.

\begin{figure}
  \includegraphics{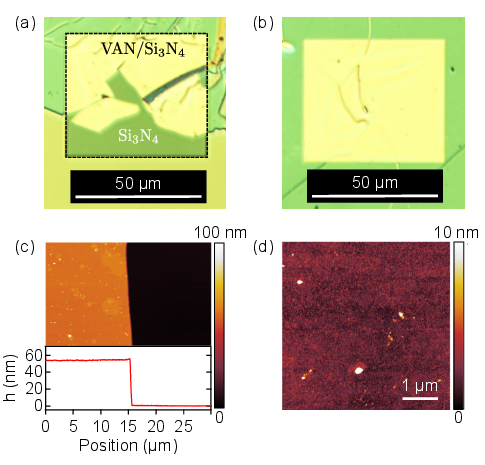}
 \caption{Delaminated membranes. (a) Optical microscopy image of a partially covered Si$_3$N$_4$ membrane window. (b) Optical microscopy image of a fully covered Si$_3$N$_4$ membrane window. (c) AFM topography scan (30 $\mu$m $\times$ 30 $\mu$m) over the edge of a flake (a profile is shown in the inset) and (d) topography scan on top of a flake (5 $\mu$m $\times$ 5 $\mu$m).}
  \label{fig:membrane}
\end{figure}

Optical microscopy images of two SN membranes are shown in Figures \ref{fig:membrane}a,b. In these images, the transferred VAN can clearly be identified. As can be seen, due to the fragility of the free-standing film, transferring large membrane parts is technically challenging. Moreover, the film may present cracks, as seen in Figure \ref{fig:membrane}a. Despite this, large micrometer-sized flakes such as the one shown in Figure \ref{fig:membrane}b, which can cover entire (50$\times$50)$\mu$m$^2$ SN membrane windows, were obtained for every sample. Note that the freestanding membrane thickness determined from atomic force microscopy (AFM) measurements (Figure \ref{fig:membrane}c) is found to be in excellent agreement with the nominal thickness of our deposited films (50\,nm). This eventually hints at the full removal of the sacrificial layer and no adhering residues after the lift-off. As can be seen in Figure 2d, on a local, micrometer scale, the surface of the membrane is free of cracks and voids and essentially flat, presenting no wrinkles and exhibiting a RMS roughness of $\approx 1$\, nm.

Having demonstrated the possibility of transferring large flakes of delaminated VAN, the next question that arises is the preservation of the integrity of the nanoarchitecture and chemical composition of the wires upon delamination. While the structural properties of the delaminated composites can be probed using conventional x-ray diffraction (as will be demonstrated in a later part of this paper), retrieving the full chemical and magnetic information is challenging, given the small diameter and large areal density of the embedded nanostructures (see the in-plane transmission electron micrograph in \textbf{SI}). To assess the properties of the wires after removal of the membrane from the substrate, we used a combination of x-ray absorption spectroscopy (XAS) and x-ray resonant magnetic scattering (XRMS) experiments. The experiments were performed at the SEXTANTS beamline of SOLEIL synchrotron, on the COMET endstation, where we took advantage of the large CCD detector (6144\,px $\times$ 6144\,px, 11$\mu$m pixel size), allowing us to access sufficiently large $q$-values in reciprocal space, corresponding to typical inter-wire distances\cite{POP19}, as will be shown in the following. 

\begin{figure}
 \includegraphics{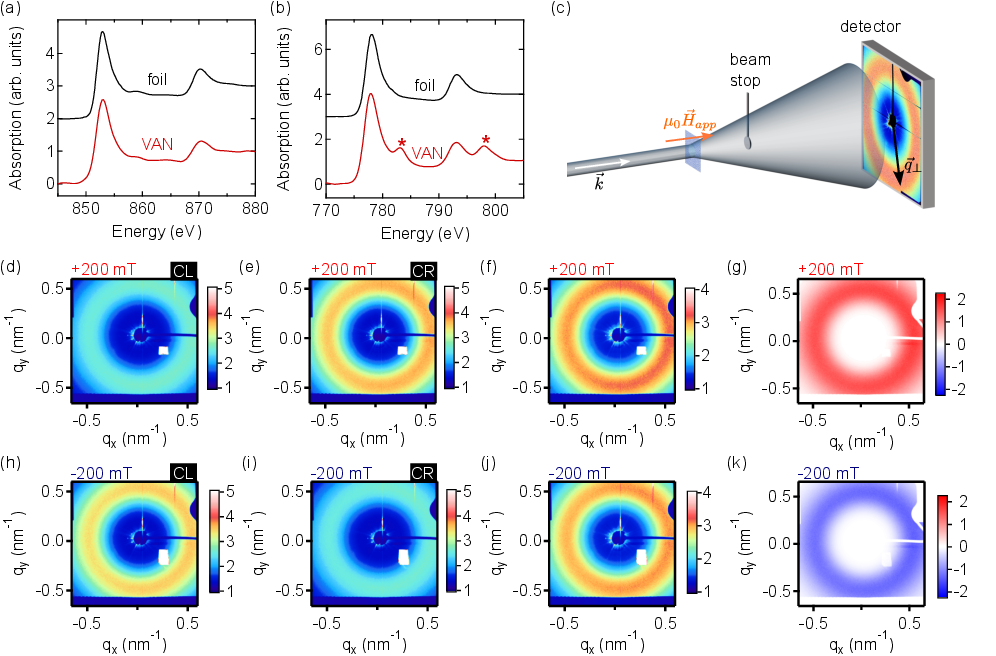}
  \caption{(a-b) X-ray absorption spectrum at the Ni (a) and Co (b) L$_{2,3}$ edges acquired in
transmission for the reference metallic foils (black) and for the delaminated VAN sample (red). Red asterisks in the Co absorption spectrum highlight the contribution from minute Ba impurities. (c) Experimental configuration for XRMS measurements on delaminated VAN samples.  (d-g) XRMS data gathered at the Ni L$_3$ edge with $\mu_0 H_{app}$=+200 mT : (d) scattering pattern acquired with circular left polarization and with (e) circular right polarization, (f) charge signal $I_c$, (g) magnetic signal $I_m$.  (h-k) XRMS data obtained by tuning the x-ray energy to the Ni L$_3$ edge with $\mu_0 H_{app}$=$-$200 mT: (h) scattering pattern acquired with circular left polarization, (i) same with a circular right polarization, (j) charge signal $I_c$, (k) magnetic signal $I_m$.  Scale bars in (d-k): intensity (arbitrary units).}
  \label{fig:XRMS}
\end{figure}

Prior to measurements on VAN membranes, the beamline energy was calibrated by using the L$_{2,3}$ edge spectra of metallic Co and Ni reference foils (using a point-like photodiode detector). The corresponding XAS spectra are shown in Figures \ref{fig:XRMS}a,b. In a second step, the same spectra were acquired for the membranes, see Figures \ref{fig:XRMS}a,b. Both the Co and Ni spectra clearly exhibit a metallic character, being nearly identical to the reference foil. No multiplet structure nor peak shift or broadening could be observed, which highlights the chemical integrity of the samples, i.e., the absence of oxide or hydroxide phases that might be induced in the nanowires by the dissolution process. Note that, in contrast to previous x-ray magnetic circular dichroism studies performed on CoNi VAN thin films on a substrate\cite{WEN22}, the spectra shown here are acquired in transmission, and not in total electron yield mode. The data are thus representative of the entire sample thickness. At the Co edge, two additional peaks are observed in the spectrum of the VAN, highlighted by red asterisks. These correspond to the X-ray absorption of Ba at the M$_{4,5}$ edges. This signal stems from minute amounts of Ba impurities in the STO matrix, which can be traced back to the PLD target used in our study\cite{WEN22}.

Scattering patterns were then acquired for the membranes using the CCD camera with x-rays tuned either to the Co or Ni L$_3$ edges. Data were gathered using consecutive measurements with opposite circular polarizations at a given magnetic field, applied in the direction perpendicular to the plane of the membrane, along the axis of the magnetic nanopillars, as illustrated in Figure \ref{fig:XRMS}c. The intensity patterns are denoted $I_+$, $I_-$ for positive and negative helicities respectively. From these measurements, two quantities are computed, following Rackham \textit{et al}. \cite{RAC23}: the average $I_c = (I_+ +I_-)/2$, which, to a good approximation, corresponds to the electron-density (charge) contribution, and the (squared) magnetic ratio $I_m = R^2_m = (I_+-I_-)^2/(I_++I_-) = I_d^2/(I_++I_-)$, which yields information about the magnetic structure of our sample. Note that while the charge scattering intensities shown in the following correspond to $I_c$, the plots of the magnetic signal correspond to $\mathrm{sign}(I_d)I_m$, which allows to highlight a possible sign switching upon reversal of the magnetization.

Exemplary scattering patterns of a Co$_{50}$Ni$_{50}$ membrane sample at the Ni L$_3$ resonance peak, acquired under a magnetic field of 200~mT (sufficiently large to saturate this specific sample) are shown in Figures \ref{fig:XRMS}d-f (see \textbf{SI} for a description of the procedure used to map the detector data into reciprocal space). The same measurements, performed with an applied field of $-$200~mT are shown in Figures \ref{fig:XRMS}h-j. Note that, as expected, the obtained annular charge patterns  $I_c(q)$ do not change when modifying the applied field (data not shown). From these data, information on the (in-plane) spatial distribution of embedded nano-objects can readily be extracted: the wires are randomly located and do not form any measurable pattern. They are however separated by a characteristic distance $\langle d\rangle$, which can be computed by fitting the intensity profiles $I_c(q)$ (obtained after azimuthal integration). This allows us to determine the position of the maximum $q_{max}$, from which $\langle d\rangle$ is calculated using $\langle d\rangle=2\pi/q_{max}$. When analysing the two transition metals separately (see \textbf{Supporting Information} for a full description of the scattering patterns obtained at the Co edge), we observe almost identical characteristic distances resulting from the charge scattering profile $I_c(q)$. On the Co$_{50}$Ni$_{50}$ sample for example, we get 12.52$\pm$0.04 nm at the Co edge, while at the Ni edge, we calculate 12.65$\pm$0.03 nm. The agreement with the periodicity value of 12.7$\pm$0.6 nm obtained from the in-plane pair-correlation function measurements by transmission electron microscopy is excellent. Identical correlation lengths for Ni and Co also indicate that the two chemical species are distributed at random in the nanowires, which is in good agreement with our earlier studies \cite{Bonilla2013, WEN22}.

In order to confirm that the magnetic properties of the nanopillars are preserved upon delamination and transfer of the membranes, we performed a similar analysis using $I_m(q)$. In contrast to the charge signal, the magnetic patterns exhibit a strong field dependence (data not shown), highlighting complex field-dependent interactions between the wires, giving rise to ordering and frustration effects (these data will be published in a follow-up paper).  Here, we exclusively focus on the magnetically saturated sample. Figures \ref{fig:XRMS}g and k show the magnetic intensity $I_m$ obtained at the Ni edge for +200 mT and at $-$200 mT. As mentioned earlier, $I_m(q)$ displays a sign reversal upon saturation in the opposite field direction. However, the location of the peak $q_{max}$ is identical for both field directions and very similar at both edges: we find $12.48 \pm 0.05$ nm using the $I_m(q)$ profile for Co and $12.59 \pm 0.04$ nm for Ni, respectively. This is in perfect agreement with the aforementioned results gathered from the charge scattering analysis and the TEM data. At saturation, all nanowires contribute to the magnetic scattering signal. This is additional evidence that allows us to rule out oxidation effects that would lead to "magnetically dead" wires (or parts of the wires), as this would have a strong impact on the magnetic correlation length. Overall, the XAS and XRMS measurements thus point towards a full preservation of the chemical and magnetic state of the nanopillars in the delamination process.

\begin{figure}
  \includegraphics{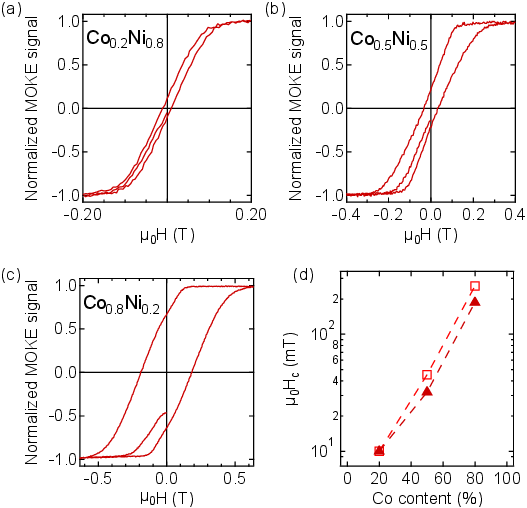}
  \caption{(a-c) Magneto-optical Kerr effect measurements of Co$_{x}$Ni$_{1-x}$-SrTiO$_3$ VAN membranes at room temperature, with the magnetic field $\mu_0 H$ applied out-of-plane, parallel to the axis of the nanopillars. (d) Coercive field as a function of the composition $x$ of Co$_{x}$Ni$_{1-x}$ nanopillars. Squares: values before delamination, triangles: values after delamination.  }
  \label{fig:MOKE}
\end{figure}

The spectroscopic analysis described in the last paragraphs was eventually complemented with MOKE measurements, performed before and after releasing the membranes from the substrate. In our nanocomposite system, the magnetic anisotropy is uniaxial and dominated by shape and magneto-elastic effects, respectively: $K_u = K_{MS} + K_{me}$. The trend observed in Figures \ref{fig:MOKE}a-c, with increasing squareness of the hysteresis cycle and coercive field indicates an increase of the anisotropy when the cobalt content increases. Despite small changes upon removal from the substrate, this trend persists in the freestanding films. Interestingly, for the Co$_{0.2}$Ni$_{0.8}$-SrTiO$_3$ VAN, $\mu_0H_c$ does not vary upon delamination (see \textbf{Supporting Information} for a full comparison of the hysteresis cycles prior to and after delamination). As shown in earlier studies, the magnetoelastic contribution vanishes for this specific composition\cite{CoNi_STRICTION}, suggesting that magnetoelastic effects, i.e., a modification of the strain state, is responsible for the observed coercivity changes.

\begin{figure}
  \includegraphics{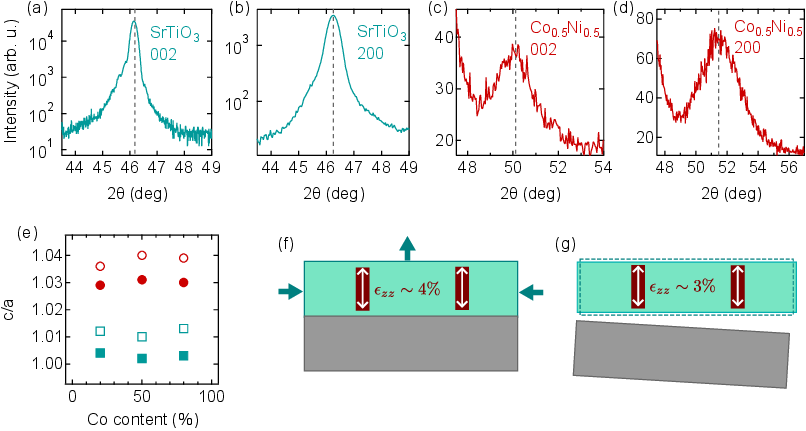}
  \caption{X-ray diffraction measurements before and after delamination and transfer of Co$_{0.5}$Ni$_{0.5}$-SrTiO$_3$ VAN on Si$_3$N$_4$ membranes. Data were collected on a laboratory 5-circle diffractometer (Rigaku SmartLab) with Cu K$ _{\alpha}$ radiation (wavelength of 1.54056~\AA ). (a-b) SrTiO$_3$ matrix: out-of-plane 002 and in-plane 200 Bragg reflections after delamination. (c-d) Co$_{0.5}$Ni$_{0.5}$ nanopillars: out-of-plane 002 and in-plane 200 Bragg reflections after delamtination. (e) $c/a$ ratios of the matrix (green) and nanopillars (red) before (open symbols) and after (plain symbols) delamination, as determined by analysis of the XRD measurements. (f-g) Schematic illustrations of the VAN strain state before (f) and after (g) delamination.}
  \label{fig:DRX}
\end{figure}

To quantify the modification of the strain state of the membranes, i.e., the dilation of the matrix and the nanowires before and after delamination, we conducted extensive x-ray diffraction measurements. In what follows, the lattice parameter along the growth axis (parallel to the axis of the nanowires) will be labeled $c$ while $a$ denotes the in-plane lattice parameters. Figure \ref{fig:DRX} gives a summary of our x-ray diffraction measurements performed on the CoNi-STO membranes (see \textbf{Supporting Information} for a full description of the out-of-plane and in-plane XRD scans). Figures \ref{fig:DRX}a-b show the 002 and 200 Bragg reflections of the SrTiO$_3$ matrix while Figures \ref{fig:DRX}c-d show the corresponding 002 and 200 Bragg reflections of the metallic alloy (Co$_{0.5}$Ni$_{0.5}$). These measurements were performed for the 3 compositions studied here, which allowed us to extract the $c/a$ ratio before and after lift-off (Figure \ref{fig:DRX}e). While the STO matrix is slightly tetragonal with $c/a$=1.010 before delamination (which is mainly due to the oxygen-defficient growth conditions that favor the formation of vacancies \cite{STOgrowth,STOgrowth2,STOgrowth3}), the metallic nanopillars exhibit a pronounced tetragonality with $c/a$ $\sim$1.04. This is less than what can be achieved in ultrathin epitaxial transition metal thin films \cite{WIN2006}. However, we emphasize that the advantage of the present composite system is that the dilation can be sustained over large distances along the wire backbone. This contrasts with planar thin films, that present a strong tendency to relax strain via defect creation once a certain critical thickness is reached. 

\vskip5mm

How does the strain in the nanocomposites evolve after delamination? Upon removal of the substrate, one would expect the elastic energy stored in the VAN layer to be released, possibly resulting in a largely distinct strain state. In the present case, we indeed observe a relaxation of the matrix. After lift-off, the tetragonality is almost completely lost, i.e., we obtain an essentially cubic STO lattice. In contrast, the tetragonality of the metallic alloy, governed by the constraints at the vertical heterointerfaces, is largely preserved with $c/a$ $\sim$1.03, irrespective of the concentration. The drop in $c/a$ of the nanowires thus follows the loss of tetragonality of the matrix, as illustrated in Figures \ref{fig:DRX} f,g. This large remanent out-of-plane dilation of the metallic Co$_x$Ni$_{1-x}$ nanostructures is one of the main findings of the present study. It demonstrates that unusual strain states can be preserved in a VAN freed from the material used for epitaxial growth. In contrast to planar systems, where the substrate removal goes along with a loss of the constraints, the delaminated nanocomposites thus remain appealing systems for strain-engineering of physical properties. In the present case, the axial strain within the ferromagnetic nanopillar is a powerful lever to act on the magnetic anisotropy, as demonstrated previously\cite{WEN22}.

In conclusion, our results demonstrate that thin metal-oxide nanocomposite films composed of vertically aligned, matrix-embedded nanowires can be fabricated via self-assembly with the help of sacrificial SrVO$_3$ layers, using intermediate ex-situ transfer steps. Even after substrate removal, their metallic state as well as their large out-of-plane dilation are preserved. This is promising for a broad range of fundamental investigations as well as for future applications: by putting VAN onto flexible, bendable supports, one obtains a versatile platform for 3D nanoscale flexomagnetism studies \cite{TANG2025}; using VAN transferred onto X-ray transparent substrates (such as the SN membranes used in the present work) allows one to use the full spectroscopy and scattering toolkit available at fourth generation light sources to investigate the properties of granular nanomagnets on ultrasmall length- and ultrafast timescales \cite{GRA2017, REP2020, TUR2022}. Finally, after further optimization of the lift-off process, one might also envision a large scale transfer of VAN membranes for integration in spintronic or optomagnetic stacks. In such heterostructures, one could harness the peculiar magnetic and transport properties of VAN, which would eventually allow to design novel types of devices that have remained out of reach so far due to the limitations imposed by top-down lithography and classical planar 2D epitaxy.

\begin{acknowledgement}
The authors acknowledge SOLEIL synchrotron for granting beam time (Proposal No. 20241560). We are grateful to the SOLEIL staff for smoothly running the facility. Part of this work is supported by Agence Nationale de la Recherche (ANR-HYPNOSE, project n$^\circ$21-CE09-0042 and ANR-FLEXO, project n$^\circ$21-CE09-0046). The authors gratefully thank N. Menguy and J.-M. Guigner, IMPMC, Sorbonne Universit\'e-CNRS, for access to the TEM facilities.
\end{acknowledgement}

\newpage

\noindent\textbf{\large{Supporting Information}} 

\renewcommand{\thefigure}{S\arabic{figure}}
\setcounter{figure}{0}

\subsection{Transmission electron microscopy}

High resolution transmission electron microscopy (HR-TEM) data were acquired using a JEOL JEM 2100F equipped with a field-emission gun operated at 200~kV.

\vskip0.5cm
\begin{figure}
  \includegraphics[scale=1.45]{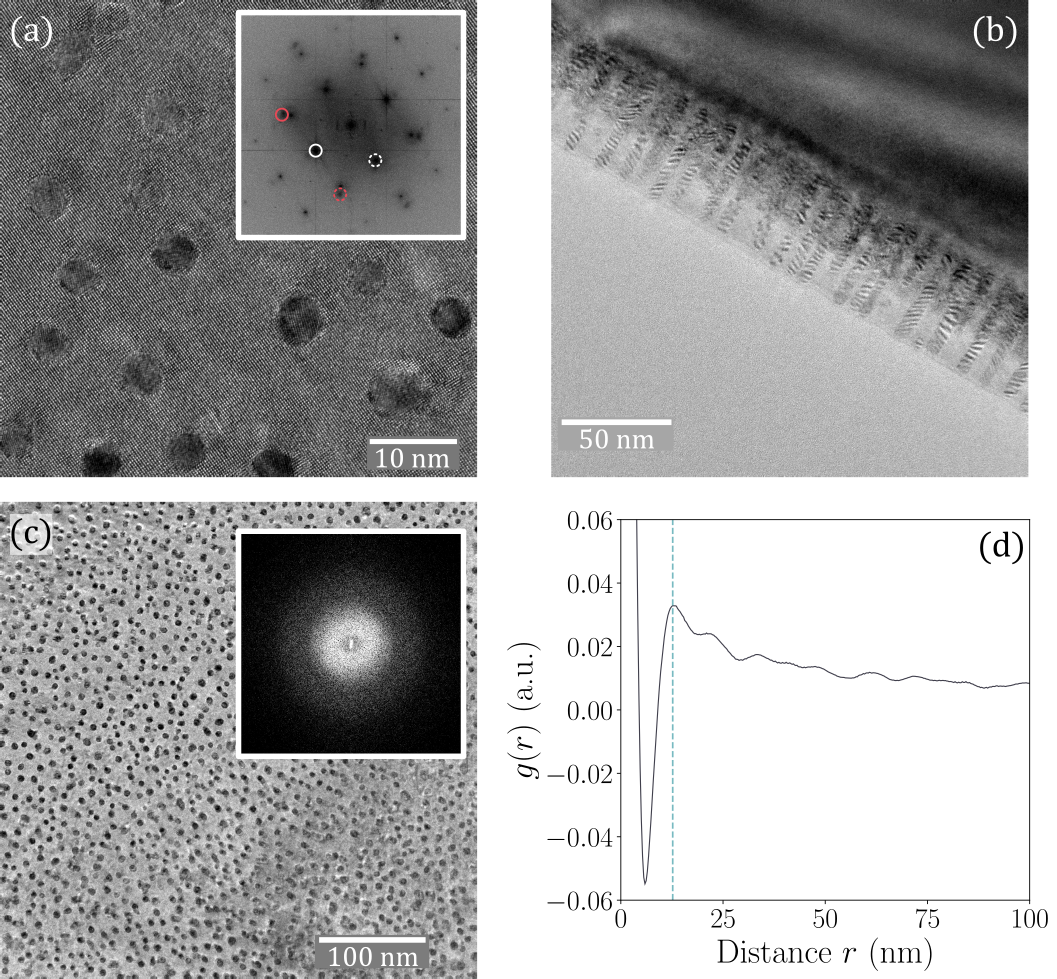}
  \vskip5mm
  \caption{High-resolution transmission electron microscopy study of a Co$_{0.5}$Ni$_{0.5}$-STO VAN. (a) High-resolution, plan-view micrograph along
the [001] zone axis (inset: fast Fourier transform of the micrograph. The full white (resp. red) circle indicates the STO 100 (resp. CoNi 1$\bar{1}$0) feature, and the dashed white (resp. red) circle highlights the STO 010 (resp. CoNi 110) feature). (b) Cross-sectional micrograph slightly tilted from the [010] zone axis. (c) Low-magnification plan-view micrograph. Inset: Fast Fourier transform of the micrograph. (d) Pair correlation function $g(r)$ of the micrograph in panel (c).}
  \label{fig:TEM_SI}
\end{figure}

\newpage
\subsection{XRMS measurements: $q$-space mapping and Co L$_{2,3}$ edge data}

To retrieve meaningful quantitative information from the scattering patterns gathered on the CCD camera, the detector data has been mapped into reciprocal space using: 
$$ q = \frac{4\pi}{\lambda} \sin\Big(\frac{1}{2} \arctan\big(\frac{ar}{D}\big)\Big)$$
Here, $\lambda$ denotes the wavelength of the x-rays, $a=11\,\mu$m the size of a pixel, $r$ the radial distance expressed in pixels and $D=20$\,cm the camera sample distance. 
\newline
\newline
X-ray resonant magnetic scattering patterns were also obtained by tuning the x-ray beam energy to the L$_3$ absorption edge of Co. All data were gathered in a transmission geometry with the $\vec{k}$-vector of the incident x-rays perpendicular to the sample surface (oriented along the nanowire axis).

\vskip0.5cm
\begin{figure}[h!]
  \includegraphics[scale=1]{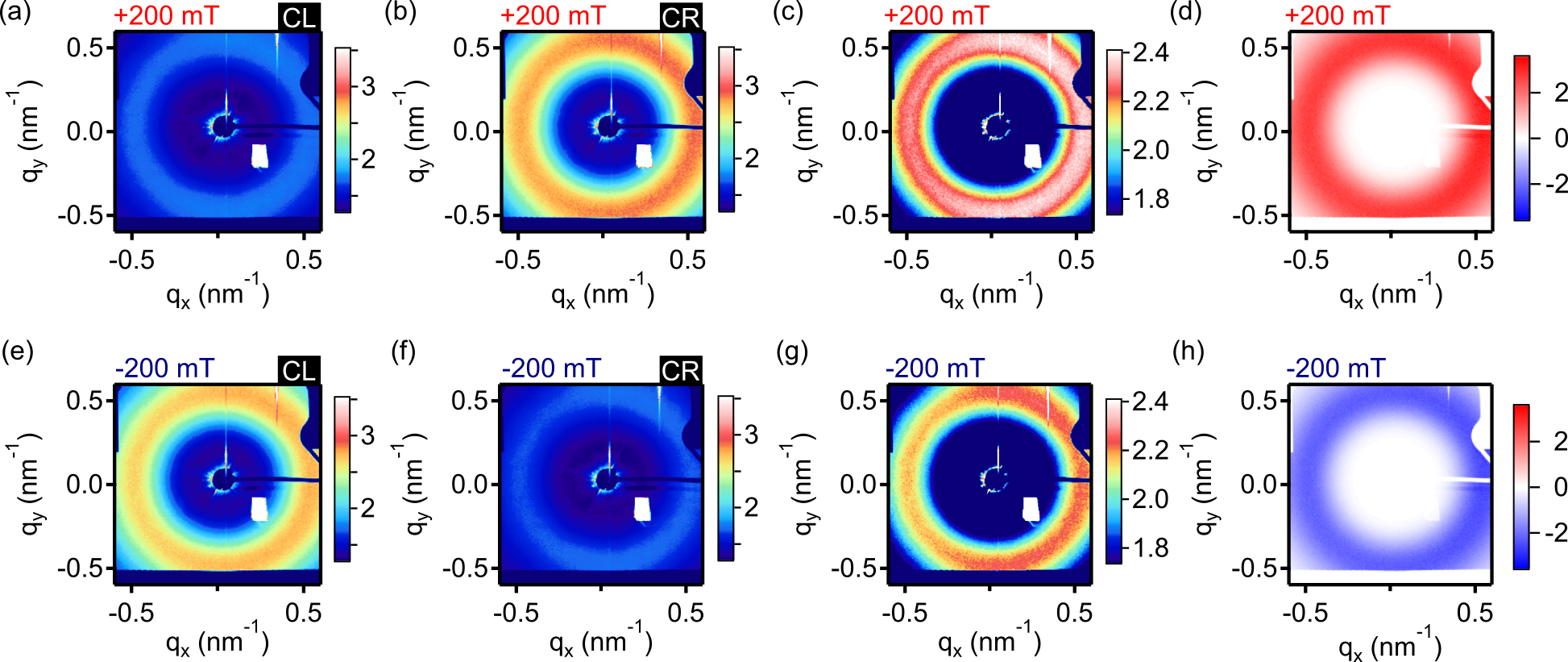}
    \vskip5mm
   \caption{XRMS data obtained at the Co edge on a Co$_{50}$Ni$_{50}$ sample. (a-d) The applied field was set to $\mu_0 H_{app}$=+200 mT : (a) scattering pattern acquired with circular left polarization, (b) same with circular right polarization, (c) charge signal $I_c$, (d) magnetic signal $I_m$.  (e-h) XRMS data at the Co edge with $\mu_0 H_{app}$=-200 mT: (e) scattering pattern acquired with circular left polarization, (f) circular right polarization, (g) charge signal $I_c$, (h) magnetic signal $I_m$ (intensity given in arbitrary units).}
  \label{fig:TEM_SI}
\end{figure}

\newpage
\subsection{Hysteresis loops $M(H)$ before and after delamination}

Magneto-optical Kerr effect measurements (P-MOKE) were performed before and after delamination on samples with composition Co$_{0.2}$Ni$_{0.8}$, Co$_{0.5}$Ni$_{0.5}$ and Co$_{0.8}$Ni$_{0.2}$. 

\vskip0.5cm
\begin{figure}
  \includegraphics[scale=1]{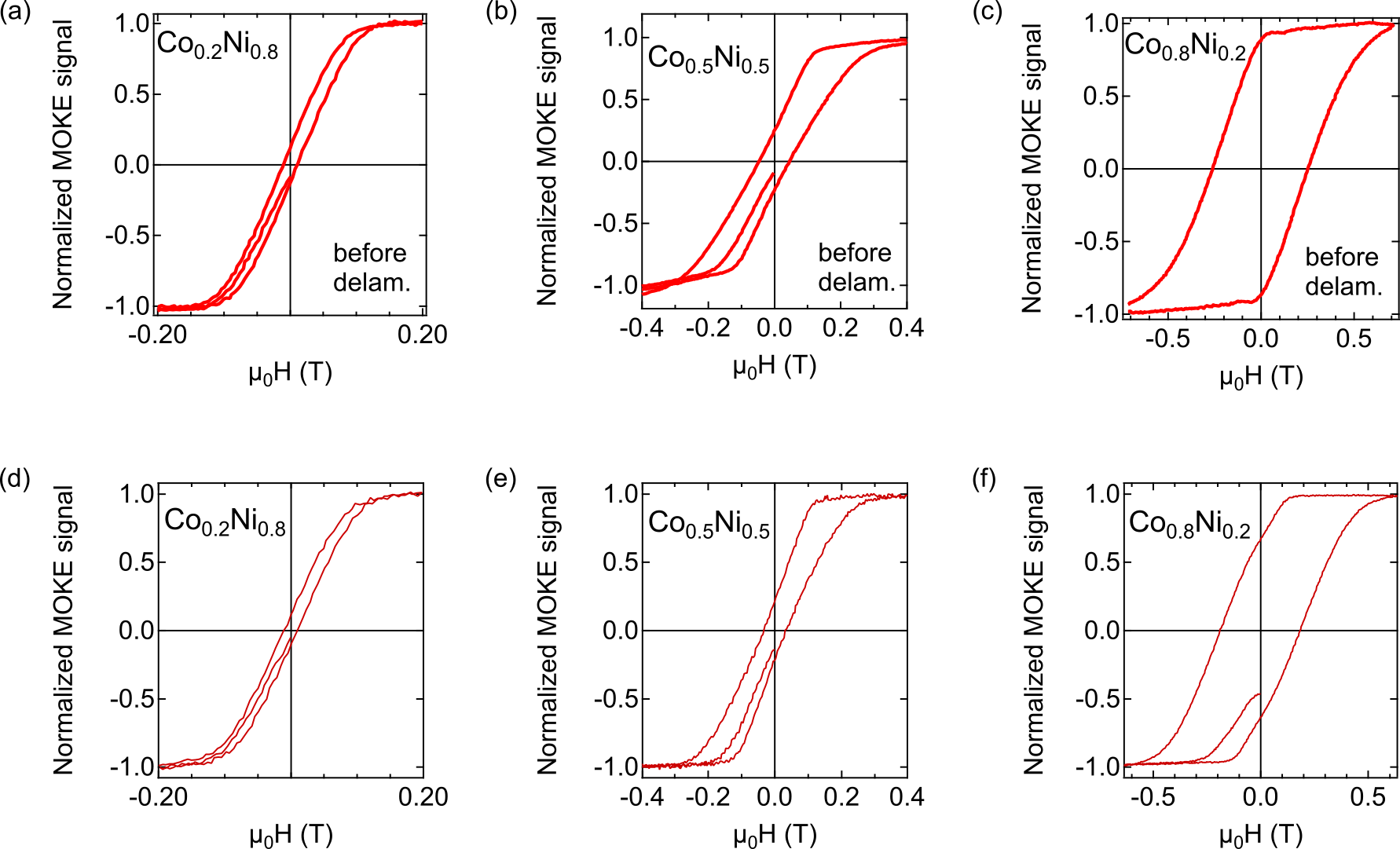}
  \caption{Normalized Kerr-rotation signal. Measurements performed before removing the membrane on (a) a Co$_{0.2}$Ni$_{0.8}$, (b) a Co$_{0.5}$Ni$_{0.5}$, and (c) a Co$_{0.8}$Ni$_{0.2}$ nanocomposite. (d-f) The same measurements were repeated on the delaminated VAN sample.}
  \label{fig:TEM_SI}
\end{figure}

\newpage
\subsection{In-plane and out-of-plane XRD scans}

X-ray diffraction data gathered on a Co$_{50}$Ni$_{50}$ VAN sample deposited on STO and after delamination and transfer to a Silson multimembrane (Si$_3$N$_4$). 

\vskip0.5cm
\begin{figure}
  \includegraphics[scale=0.85]{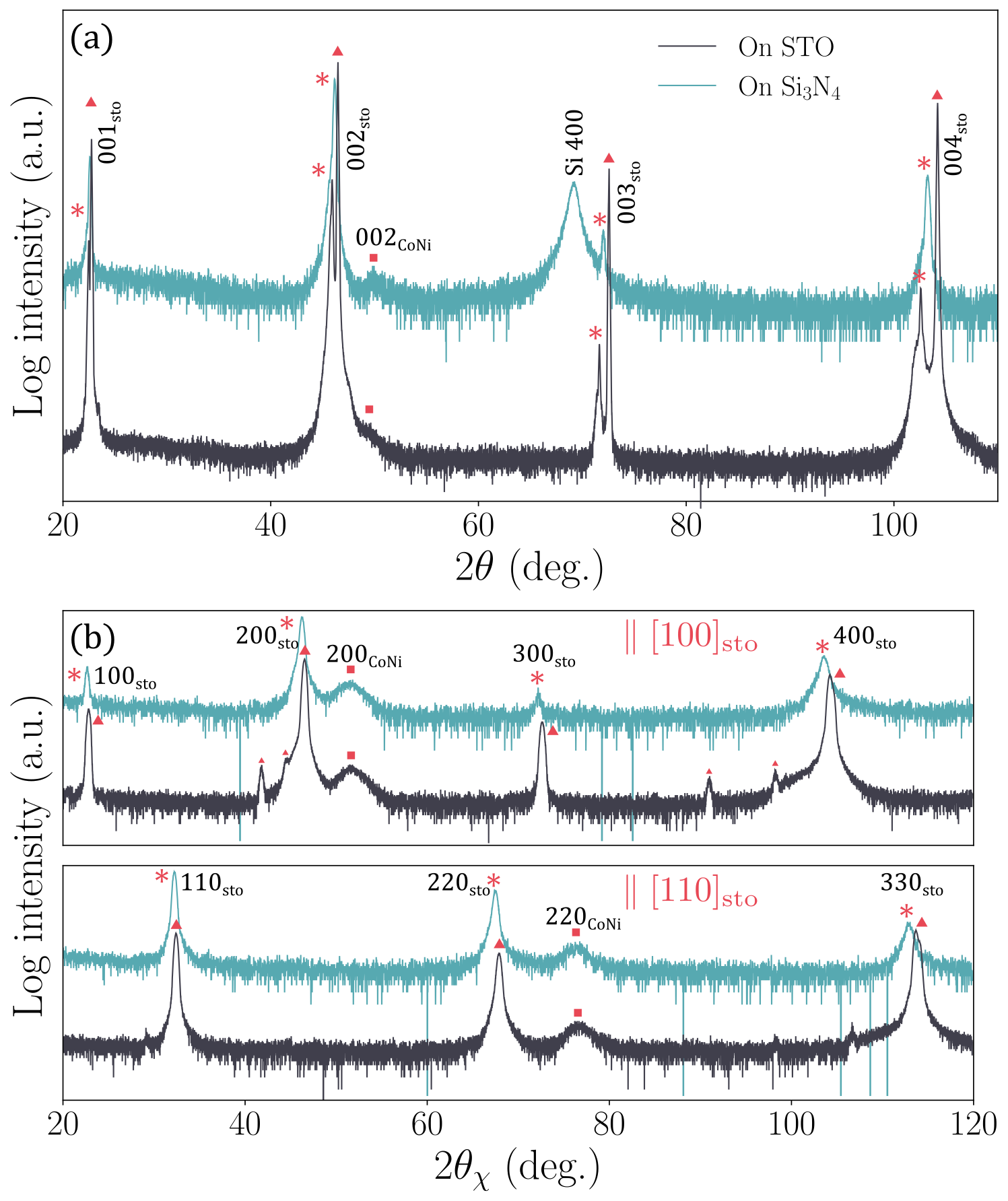}
  \caption{X-ray diffraction measurements. X-ray crystallography of a CoNi-STO VAN prior to and after delamination. (a) $\theta$–2$\theta$ pattern (out-of-plane measurement) before (black) and after transfer (blue). (b) In-plane 2$\theta_{\chi}$–$\Phi$ scans along the [100]STO and the [110]STO directions, before (black) and after transfer (blue). Red triangles: STO substrate peaks, Red stars: STO matrix peaks, red squares: CoNi peaks. (Smaller triangles correspond to W L$\alpha_1$,  W L$\alpha_2$ and Cu K$\beta$ radiation, since no monochromator was used for IP measurements.)}
  \label{fig:TEM_SI}
\end{figure}


\begin{thebibliography}{2}

\bibitem{ZHA19} Zhang, Y.; Ma, C.; Lu, X.; Liu, M. Recent progress on flexible inorganic single-crystalline functional oxide films for advanced electronics. \textit{Mater. Horiz.} \textbf{2019}, \textit{6}, 911.

\bibitem{PESQ22} Pesquera, D.; Fern{\'a}ndez, A; Khestanova, E; and Martin, L. W., Freestanding complex-oxide membranes. \textit{Journal of Physics: Condensed Matter} \textbf{2022} \textit{34}, 383001 

\bibitem{CHIA22} Chiabrera, F. M.; Yun, S.; Li, Y.; Dahm, R. T.; Zhang, H.; Kirchert, C. K. R.; Christensen, D. V.; Trier, F.; Jespersen, T. S.; Pryds, N., Freestanding perovskite oxide films: Synthesis, challenges, and properties, \textit{Annalen der Physik},
  \textbf{2022}, \textit{534} 2200084.

\bibitem{MAN25} Mandal, R.; Yun, S.; Wurster, K.; Dollekamp, E.; Shondo, J. N.; Pryds, N. Recent Advancement in Ferroic Freestanding Oxide Nanomembranes. \textit{Nano Lett.} \textbf{2025}, \textit{25}, 5541 -- 5549.

\bibitem{Ji2019} Ji, D.; Cai, S.; Paudel, T.R.; Sun, H.; Zhang, C.; Han, L.; Wei, Y.; Zang, Y.; Gu, M.; Zhang, Y.; Gao, W.; Huyan, H.; Guo, W.; Wu, D.; Gu, Z.; Tsymbal, E. Y.; Wang, P.; Nie, Y.; Pan, X. Freestanding crystalline oxide perovskites down to the monolayer limit. \textit{Nature} \textbf{2019}, \textit{570}, 87--90.

\bibitem{lu2016synthesis} Lu, D.; Baek, D. J.; Hong, S. S.; Kourkoutis, L. F.; Hikita, Y.; Hwang, H. Y. Synthesis of freestanding single-crystal perovskite films and heterostructures by etching of sacrificial water-soluble layers \textit{Nat. Mater.} \textbf{2016}, \textit{15} 1255--1260

\bibitem{BOU20} Bourlier, Y.; B\'erini, B.; Fr\'egnaux, M.; Fouchet, A.; Aureau, D.; Dumont, Y. Transfer of Epitaxial SrTiO$_3$ Nanothick Layers Using Water-Soluble Sacrificial Perovskite Oxides. \textit{ACS Appl. Mater. Interfaces} \textbf{2020}, \textit{12}, 8466--8474. 

\bibitem{TAKA2020} Takahashi, R.; Lippmaa, M. Sacrificial water-soluble BaO layer for fabricating free-standing piezoelectric membranes. \textit{ACS Appl. Mater. Interfaces} \textbf{2020}, \textit{12}, 25042--25049.

\bibitem{GAN24} Ganguly, S.; Pesquera, D.; Garcia, M.; Saeed, U.; Mirzamohammadi, N.; Santiso, J.; Padilla, J.; Roque, J. M. C.; Laulhé, C.; Berenguer, F.; Villanueva, L. G.; Catalan, G. Photostrictive Actuators Based on Freestanding Ferroelectric Membranes. \textit{Adv. Mater.} \textbf{2024}, \textit{36}, 2310198. 

\bibitem{HUA22} Huang, J. K.; Wan, Y.; Shi, J.; Zhang, J.; Wang, Z.; Wang, W.; Yang, N.; Liu, Y.; Lin, C. H.; Guan, X.; Hu, L.; Yang, Z. L.; Huang, B. C.; Chiu, Y. P.; Yang, J.; Tung, V.; Wang, D.; Kalantar-Zadeh, K.; Wu, T.; Zu, X.; Qiao, L.; Li, L. J.; Li, S. High-$\kappa$ perovskite membranes as insulators for two-dimensional transistors. \textit{Nature} \textbf{2022}, \textit{605}, 262--267.

\bibitem{SU23} Su, Y.; Zong, A.; Kogar, A.; Lu, D.; Hong, S. S.; Freelon, B.; Rohwer, T.; Wang, B. Y.; Hwang, H. Y.; Gedik, N. Delamination-Assisted Ultrafast Wrinkle Formation in a Freestanding Film, \textit{Nano Lett.} \textbf{2023}, \textit{23}, 10772-- 10778.

\bibitem{HAR23} Harbola, V.; Wu, Y.-J.; Wang, H.; Smink, S.; Parks, S. C.; van Aken, P. A.; Mannhart J. Self-Assembly of Nanocrystalline Structures from Freestanding Oxide Membranes. \textit{Adv. Mater.} \textbf{2023}, \textit{35}, 2210989.

\bibitem{SEG24} Segantini, G.; Hsu, C.-Y.; Rischau, C. W.; Blah, P.; Matthiesen, M.; Gariglio, S.; Triscone, J.-M.; Alexander, D. T. L.; Caviglia, A. D.  Electron-Beam Writing of Atomic-Scale Reconstructions at Oxide Interfaces. \textit{Nano Lett.} \textbf{2024}, \textit{24}, 14191 --14197

\bibitem{DEG25} Degezelle, A.; Burcea, R.; Gemeiner, P.; Vallet, M.; Dkhil, B.; Fusil, S.; Garcia, V.; Matzen, S.; Lecoeur, P.; Maroutian, T.  Strain-Induced Polarization Rotation in Freestanding Ferroelectric Oxide Membranes. \textit{Adv. Electron. Mater.} \textbf{2025}, e00266. 

\bibitem{SHE23} Sheeraz, M.; Jung, M.-H.; Kim, Y. K.; Lee, N.-J.; Jeong, S.; Choi, J. S.; Jo,Y. J.; Cho, S.; Kim, I. W.; Kim, Y.-M.; Kim, S.; Ahn, C. W.; Yang, S. M.; Jeong, H. Y.; Kim, T. H. Freestanding Oxide Membranes for Epitaxial Ferroelectric Heterojunctions. \textit{ACS Nano}  \textbf{2023}, \textit{17}, 13510 -- 13521.

\bibitem{Macmanus2010} MacManus-Driscoll, J. L. Self-Assembled Heteroepitaxial Oxide Nanocomposite Thin Film Structures: Designing Interface-Induced Functionality in Electronic Materials. \textit{Adv. Func. Mat.} \textbf{2010}, \textit{20}, 2035--2045.

\bibitem{Zheng2014} Zhang, W.; Chen, A.; Bi, Z.; Jia, Q.; MacManus-Driscoll, J. L.; Wang, H. Interfacial Coupling in Heteroepitaxial Vertically Aligned Nanocomposite Thin Films: From Lateral to Vertical Control. \textit{Curr. Opin. Solid State Mater. Sci.} \textbf{2014}, \textit{18}, 6--18.

\bibitem{Review2018}  Chen, A.; Su, Q.; Han, H.; Enriquez, E.; Jia, Q. Metal Oxide Nanocomposites: A Perspective from Strain, Defect, and Interface. \textit{Adv. Mater.} \textbf{2018}, \textit{31}, 1803241.

\bibitem{LIP2016} Kawasaki, S.; Takahashi, R.; Yamamoto, T.; Kobayashi, M.; Kumigashira, H.; Yoshinobu, J.; Komori, F.; Kudo, A.; Lippmaa, M. Photoelectrochemical water splitting enhanced by self-assembled metal nanopillars embedded in an oxide semiconductor photoelectrode. \textit{Nat. Commun.} \textbf{2016}, \textit{7}, 11818.

\bibitem{LIP2021} Lippmaa, M.; Kawasaki, S.; Takahashi, R.; Yamamoto, T. Nanopillar composite electrodes for solar-driven water splitting. \textit{MRS Bulletin} \textbf{2021}, \textit{46}, 142--151.

\bibitem{CHEN2013} Chen, A.; Bi, Z.; Jia, Q.; MacManus-Driscoll, J. L.;  Wang, H. Microstructure, vertical strain control and tunable functionalities in self-assembled, vertically aligned nanocomposite thin films. \textit{Acta Mater.} \textbf{2013}, \textit{61}, 2783--2792.

\bibitem{WAN2016}  Li, L.; Sun, L.; Gomez-Diaz, J. S.; Hogan, N. L.; Lu, P.; Khatkhatay, F.; Zhang, W.; Jian, J.; Huang, J.; Su, Q.; Fan, M.; Jacob, C.; Li, J.; Zhang, X.; Jia, Q.; Sheldon, M.; Al\`u, A.; Li, X.;  Wang, H. Self-assembled epitaxial Au–Oxide vertically aligned nanocomposites for nanoscale metamaterials. \textit{Nano Lett.} \textbf{2016}, \textit{16}, 3936--3943.
   
\bibitem{WU2024} Wu, J.; Yan, F.; Zhao, J.; Qian, L.; Cheng, T.-H.; Su, J.; Bi, L.; Huang, Y.; Wang, W.; Zhang, Z.; Luo, F.; Ning, S. Dynamically Tunable Localized Surface Plasmon Resonance in Self‐Assembled SrCoO$_x$‐Au Vertically Aligned Nanocomposite Thin Films. \textit{Adv. Funct. Mater.} \textbf{2024} \textit{34} 2411358.

\bibitem{GAO2021} Gao, M.; Yang, Y.; Rao, W.-F.; Viehland, D. Magnetoelectricity in vertically aligned nanocomposites: Past, present, and future. \textit{MRS Bulletin} \textbf{2021}, \textit{46}, 123--130. 

\bibitem{ROSS2015} Aimon, N. M.; Kim, D. H.; Sun, X. Y.; Ross, C. A. Multiferroic behavior of templated BiFeO$_3$–CoFe$_2$O$_4$ self-assembled nanocomposites. \textit{ACS Appl. Mater. Interfaces} \textbf{2015}, \textit{7}, 2263--2268.

\bibitem{DOU2023} Dou, H.; Lin, Z.; Hu, Z.; Tsai, B. K.; Zheng, D.; Song, J.; Lu, J.; Zhang, X.; Jia, Q.; MacManus-Driscoll, J. L.; Ye, P. D.; Wang, H. Self-assembled Au nanoelectrodes: enabling low-threshold-voltage HfO$_2$-based artificial neurons. \textit{Nano Lett.} \textbf{2023}, \textit{23}, 9711--9718.

\bibitem{CHE21} Chen A.; Quanxi Jia, Q. A pathway to desired functionalities in vertically aligned nanocomposites and related architectures. \textit{MRS Bulletin} \textbf{2021}, \textit{46}, 115 (2021).

\bibitem{HEN21} Hennes, M.; Demaille, D.; Patriarche, G.; Tran, T.; Zheng, Y.; Vidal, F. Strain, magnetic anisotropy, and composition modulation in hybrid metal-oxide vertically assembled nanocomposites. \textit{MRS Bulletin} \textbf{2021}, \textit{46}, 136 (2021).

\bibitem{LIU2012} Liu, H.-J.; Chen, L.-Y.; He, Q.; Liang, C.-W.; Chen, Y.-Z.; Chien, Y.-S.; Hsieh, Y.-H.; Lin, S.-J.; Arenholz, E.; Luo, C.-W.; Chueh, Y.-L.; Chen, Y.-C.; Chu, Y.-H. Epitaxial photostriction–magnetostriction coupled self-assembled nanostructures.\textit{ACS Nano} \textbf{2012}, \textit{6} 6952--6959.

\bibitem{Huang2025}{Huang, J.; Tsai, B. K.; Choudhury, A.; Shen, J.; Mihalko, C. A.; Zhou, S.; Liu, C.; Wang, H. Freestanding TiN‐Au Vertically Aligned Nanocomposite Thin Films for Flexible Plasmonic Hybrid Metasurfaces. \textit{Adv. Mater. Interfaces} \textbf{2025} e00613.}

\bibitem{TSAI24} Tsai, B. K.; Huang, J.; Yu, Y.-C.; Lee, M. H.; Stegman, B. T.; Flores, E. J.; Tong, P. Z.; Xu, K.; Zhou, S.; Shen,  J.; Song, J.; Zhang, Y.; Stanciu, L.; Wu, W.; Zhang, X.; Wang, H. Freestanding BaTiO$_3$-Au Vertically Aligned Nanocomposite toward Flexible Multi-Sensing Platform. \textit{Adv. Funct. Mater.}  \textbf{2025}, 2418004

\bibitem{MOH04} Mohaddes-Ardabili, L.; Zheng, H.; Ogale, S. B.; Hannoyer, B.; Tian, W.; Wang, J.; Lofland, S. E.; Shinde, S. R.; Zhao, T.; Jia, Y.; Salamanca-Riba, L.; Schlom, D. G.; Wuttig, M.; Ramesh, R. Self-assembled single-crystal ferromagnetic iron nanowires formed by decomposition. \textit{Nat. mater.}, \textbf{2004}, \textit{3}, 533-538.

\bibitem{SHI12} Shin, J.; Goyal, A.; Cantoni, C.; Sinclair, J. W.; Thompson J. R. Self-assembled ferromagnetic cobalt/yttria-stabilized zirconia nanocomposites for ultrahigh density storage applications. \textit{Nanotechnology} \textbf{2012} \textit{23}, 155602.

\bibitem{Bonilla2013} Bonilla, F. J.; Novikova, A.; Vidal, F.; Zheng, Y. L.; Fonda, E.; Demaille, D.; Schuler, V.; Coati, A.; Vlad, A.; Garreau, Y.; Sauvage Simkin, M.; Dumont, Y.; Hidki, S.; Etgens, V. Combinatorial Growth and Anisotropy Control of Self-Assembled Epitaxial Ultrathin Alloy Nanowires. \textit{ACS Nano} \textbf{2013}, \textit{7}, 4022--4029.

\bibitem{SCH16} Schuler, V.; Milano, J.; Coati, A.; Vlad, A.; Sauvage-Simkin, M.; Garreau, Y.; Demaille, D.; Hidki, S.; Novikova, A.; Fonda, E.; Zheng, Y.; Vidal, F. Growth and magnetic properties of vertically aligned epitaxial CoNi nanowires in (Sr, Ba)TiO$_3$ with diameters in the 1.8–6 nm range. \textit{Nanotechnology}, \textbf{2016}, \textit{27}, 495601.

\bibitem{SU16} Su, Q.; Zhang, W.; Lu, P.; Fang, S.; Khatkhatay, F.; Jian, J.;  Li, L.; Chen, F.; Zhang, X.; MacManus-Driscoll, J. L.; Chen, A.; Jia, Q.; Wang, H. Self-Assembled Magnetic Metallic Nanopillars in Ceramic Matrix with Anisotropic Magnetic and Electrical Transport Properties. \textit{ACS Appl. Mater. Interfaces}, \textbf{2016}, \textit{8}, 20283-20291.


\bibitem{HUA17} Huang, J.; Li, L.; Lu, P.; Qi, Z.; Sun, X.; Zhang, X.; Wang, H. Self-assembled Co–BaZrO$_3$ nanocomposite thin films with ultra-fine vertically aligned Co nanopillars. \textit{Nanoscale}, \textbf{2017}, \textit{9}, 7970-7976.

\bibitem{HEN18} Hennes, M.; Schuler, V.; Weng, X.; Buchwald, J.; Demaille, D.; Zheng, Y.; Vidal, F. Growth of vertically aligned nanowires in metal–oxide nanocomposites: kinetic Monte-Carlo modeling versus experiments. \textit{Nanoscale}, \textbf{2018} \textit{10}, 7666-7675.

\bibitem{WEN18} Weng, X.; Hennes, M.; Coati, A.; Vlad, A.; Garreau, Y.; Sauvage-Simkin, M; Fonda, E.; Patriarche, G.; Demaille, D.; Vidal, F.; Zheng, Y. Ultrathin Ni nanowires embedded in SrTiO$_3$: Vertical epitaxy, strain relaxation mechanisms, and solid-state amorphization. \textit{Phys. Rev. Mater.} \textbf{2018}, \textit{2}, 106003.

\bibitem{WEN22} Weng, X.; Hennes, M.; Juhin, A.; Sainctavit, Ph.; Gobaut, B.; Otero, E.; Choueikani, F.; Ohresser, P.; Tran, T.; Hrabovsky, D.; Demaille, D.; Zheng, Y.; Vidal, F. Strain-engineering of magnetic anisotropy in Co$_x$Ni$_{1-x}$-SrTiO$_3$/SrTiO$_3$(001) vertically assembled nanocomposites. \textit{Phys. Rev. Mater.} \textbf{2022}, \textit{6}, 046001.

\bibitem{HUA2024} Huang, J. Li, L.; Hu, Z.; Tsai, B. K.; Huang, J.; Shen, J.; Zhang, Y.; Barnard, J. P.; Song, J.; Wang, H. Ultrathin Ternary FeCoNi Alloy Nanoarrays in BaTiO$_3$ Matrix for Room-Temperature Multiferroic and Hyperbolic Metamaterial. \textit{Nano Lett.} \textbf{2024}, \textit{24} 10081--10089.

\bibitem{FOU16} Fouchet, A.; Allain, M.; B{\'e}rini, B.; Popova, E.; Janolin, P.-E.; Guiblin, N.; Chikoidze, E.; Scola, J.; Hrabovsky, D.; Dumont, Y. Study of the electronic phase transition with low dimensionality in SrVO$_3$ thin films. \textit{Materials Science and Engineering: B}, \textbf{2016}, \textit{212} 7--13.

\bibitem{POP19} H. Popescu, H.; Perron, J.; Pilette, B.; Vacheresse, R.; Pinty, V.; Gaudemer, R; Sacchi, M.; Delaunay, R.; Fortuna, F.; Medjoubi, K.; Desjardins, K.; Luning, J.; Jaouen, N. COMET: a new end-station at SOLEIL for coherent magnetic scattering in transmission, \textit{J. Synchrotron Radiat.} \textbf{2019} \textit{26}, 280.

\bibitem{RAC23} Rackham, J.; Pratt, B.; Griner, D.; Smith, D.; Cai, Y.; Harrison, R. G.; Reid, A.; Kortright, J.; Transtrum, M. K.; Chesnel, K. Field-dependent nanospin ordering in monolayers of Fe$_3$O$_4$ nanoparticles throughout the superparamagnetic blocking transition, \textit{Phys. Rev. B} \textbf{2023}, \textit{108}, 104415.

\bibitem{STOgrowth} Suzuki, T.; Nishi, Y.; Fujimoto, M. Defect structure in homoepitaxial non-stoichiometric strontium titanate thin films. \textit{Philos. Mag. A} \textbf{2000}, \textit{80}, 621.

\bibitem{STOgrowth2} Ohnishi, T.; Shibuya, K.; Yamamoto, T.; Lippmaa, M. Defects and transport in complex oxide thin films. \textit{J. Appl. Phys.} \textbf{2008}, \textit{103}, 103703.

\bibitem{STOgrowth3} Brooks, C. M.;  Kourkoutis, L. F.; Heeg, T.; Schubert, J.; Muller, D. A.; Schlom, D. G. Growth of homoepitaxial SrTiO$_3$ thin films by molecular-beam epitaxy. \textit{Appl. Phys. Lett.} \textbf{2009}, \textit{94}, 162905.

\bibitem{CoNi_STRICTION} Kadowaki, S.; Takahashi, M. Magnetostriction Constants of Nickel-Cobalt Alloys  \textit{J. Phys. Soc. Jap.}, \textbf{1981}, \textit{50}, 1154--1161.

\bibitem{WIN2006} Winkelmann, A.; Przybylski, M.; Luo, F.; Shi, Y.; Barthel, J. Perpendicular Magnetic Anisotropy Induced by Tetragonal Distortion of FeCo Alloy Films Grown on Pd(001). \textit{Phys. Rev. Lett.} \textbf{2006}, \textit{96} 257205.

\bibitem{TANG2025} Tang, Z.; Gong, Q.;  Yi, M. Flexomagnetism: Progress, challenges, and opportunities. \textit{Mater. Sci. Eng. R Rep.}, \textbf{2025}, \textit{62}, 100878.

\bibitem{GRA2017} Granitzka, P. W.; Jal, E.; Le Guyader, L.; Savoini, M.; Higley, D. J.; Liu, T.; Chen, Z.; Chase, T.; Ohldag, H.; Dakovski, G. L.; Schlotter, W. F.; Carron, S.; Hoffman, M. C.; Gray, A. X.; Shafer, P.; Arenholz, E.; Hellwig, O.; Mehta, V.; Takahashi, Y. K.; Wang, J.; Fullerton, E. E.; St\"ohr, J.; Reid, A. H.; D\"urr, H. A. Magnetic Switching in Granular FePt Layers Promoted by Near-Field Laser Enhancement. \textit{Nano. Lett.} \textbf{2017}, \textit{17}, 2426--2432.

\bibitem{REP2020} Von Reppert, A.; Willig, L.; Pudell, J.-E.; Zeuschner, S. P.; Sellge, G.; Ganss, F.; Hellwig, O.; Arregi, J. A.; Uhlíř, V.; Crut, A.; Bargheer, M., Spin stress contribution to the lattice dynamics of FePt, \textit{Sci. Adv.} \textbf{2020}, \textit{6}, eaba1142. 

\bibitem{TUR2022} Turenne, D.; Yaroslavtsev, A.; Wang, X.; Unikandanuni, V.; Vaskivskyi, I.; Schneider, M.; Jal, E.; Carley, R.; Mercurio, G.; Gort, R.; Agarwal, N.; Van Kuiken, B.; Mercadier, L.; Schlappa, J.; Le Guyader, L.; Gerasimova, N.; Teichmann, M.; Lomidze, D.; Castoldi, A.; Potorochin, D.; Mukkattukavil, D.; Brock, J.; Zhou Hagström, N.; Reid A. H.; Shen, X.; Wang, X. J.; Maldonado, P.; Kvashnin, Y.; Carva, K.; Wang, J.; Takahashi, Y. K.; Fullerton, E. E.; Eisebitt, S.; Oppeneer, P. M.; Molodtsov, S.; Scherz, A.; Bonetti, S.; Iacocca, E.; Dürr, H. A., Nonequilibrium sub–10 nm spin-wave soliton formation in FePt nanoparticles, \textit{Sci. Adv.}, \textbf{2022}, \textit{8}, eabn0523.

\end{thebibliography}
\end{document}